\documentclass[letterpaper, 10 pt, conference]{ieeeconf}  

\IEEEoverridecommandlockouts                              

\usepackage{graphics} 
\graphicspath{{./figures/}}
\usepackage{subcaption}
\usepackage{amsmath} 

\usepackage{amsthm}
\usepackage{amssymb}  
\usepackage{booktabs}
\usepackage{amsmath,amsfonts,amssymb,amscd} 
\usepackage{mathtools,mathrsfs,bm,amsthm}
\newcommand{\colvec}[2][.9]{%
  \scalebox{#1}{%
    \renewcommand{\arraystretch}{1}%
    $\begin{bmatrix}#2\end{bmatrix}$%
  }
}
\usepackage{braket}
\usepackage{dsfont}
\usepackage{accents}
\usepackage{siunitx}
\usepackage{cite}
\usepackage{multirow}
\usepackage{booktabs}
\usepackage{url}
\usepackage{siunitx}

\usepackage[linesnumbered, ruled]{algorithm2e}
\SetKwComment{Comment}{/* }{ */}
\SetKwRepeat{Do}{do}{while}

\usepackage{tikz}
\usetikzlibrary{patterns,decorations.pathreplacing,decorations.markings,backgrounds,calc,intersections,arrows,arrows.meta,fit,positioning,3d}
\usepackage{tikz-3dplot}

\makeatletter
\let\NAT@parse\undefined
\makeatother

\usepackage[hidelinks]{hyperref}

\renewcommand{\vec}[1]{\bm{\mathbf{#1}}}

\newcommand{\real}{\mathbb{R}}

\DeclareMathOperator{\diag}{diag}
\DeclareMathOperator*{\argmin}{argmin}
\DeclareMathOperator{\image}{Im}

\DeclarePairedDelimiter{\norm}{\lVert}{\rVert}

\theoremstyle{definition}

\theoremstyle{remark}
\newtheorem{remark}{Remark}

\usepackage[
placement=top,
angle=0,
color=red,
scale=1.1,
hshift=0,
vshift=-15
]{background}
\backgroundsetup{contents=\bf{----------------------------------------------------------------- PREPRINT -----------------------------------------------------------------}}

\title{\LARGE \bf
An Active-Sensing Approach for Bearing-based Target Localization
}

\author{Beniamino Pozzan$^{1}$, Giulia Michieletto$^{2}$, Mehran Mesbahi$^{3}$, Angelo Cenedese$^{1}$
\thanks{This work is partially supported by MUR through PRIN Grant DOCEAT 2020RTWES4 and the Gini Foundation.}
\thanks{$^{1}$ Beniamino Pozzan and Angelo Cenedese are with the Department of Information Engineering, University of Padova, Italy
        }%
\thanks{$^{2}$ Giulia Michieletto is with the Department of Management and Engineering, University of Padova, Italy
        }%
\thanks{$^{3}$ Mehran Mesbahi is with the William E. Boeing Department of Aeronautics and Astronautics, University of Washington, Seattle, WA, USA
        }%
\thanks{Corresponding contact:
{\tt\small giulia.michieletto@unipd.it}
        }
}

\begin{document}

\maketitle
\thispagestyle{empty}
\pagestyle{empty}

\begin{abstract}

Characterized by a cross-disciplinary nature, the bearing-based target localization task involves estimating the position of an entity of interest by a group of agents capable of collecting noisy bearing measurements. In this work, this problem is tackled by resting both on the weighted least square estimation approach and on the active-sensing control paradigm. Indeed, we propose an iterative algorithm that provides an estimate of the target position under the assumption of Gaussian noise distribution, which can be considered valid when more specific information is missing. Then, we present a seeker agents control law that aims at minimizing the localization uncertainty by optimizing the covariance matrix associated with the estimated target position. 
The validity of the designed bearing-based target localization solution is confirmed by the results of an extensive Monte Carlo simulation campaign.
\end{abstract}

\section{INTRODUCTION}
\label{sec:intro}

The target localization task involves the estimation of the position of an object, a person, or an event, based on the data collected by some sensing agents while surveying a specific area of interest.
This issue spans multiple application scenarios, including environmental monitoring (e.g., radiation sources identification, forest fires detection), territorial surveillance (e.g., intruders/potential threats recognition), and search-and-rescue missions (e.g., localization of victims of catastrophic events)~\cite{robin2016multi}.\\
Due to its widespread occurrence, target localization remains an intriguing research topic within the field of robotics and control.
Current solving methods vary depending on several application factors such as the number of target and seeker agents, their ability to act (distinguishing between static and dynamic agents), and the type of involved sensing data, including bearing and/or distance measurements.\\
In this work, the attention is focused on the localization of a single static target by a group of dynamic seeker agents having bearing sensing capabilities.
In particular, the task is addressed by accounting for the twofold aspect of the target position estimate and the seeker agents control.
\medskip 

\noindent{\textit{Related works}} - 
The problem of determining the unknown position of a certain object by exploiting a given set of bearing measurements dates back to~\cite{stansfield1947statistical} where the expected root-mean-square error on the target position estimate is adopted as the index for measuring the estimation reliability.
Later, multiple works cope with the bearing-based target localization task. Without claiming to be exhaustive, we mention~\cite{kaplan2001maximum} and~\cite{bishop2009bearing}, which specifically focus on the measurements features.
Indeed, in the former, a Maximum Likelihood approach is exploited to face the non-linearities affecting the measurement noise, while in the latter the task is formalized and addressed in a constrained geometric optimization framework resting on the relative direction of the agents' bearings.
More recently, similar measurement-aware methods have been proposed for the localization of multiple static targets by means of a multi-agent formation in the context of obstacle avoidance task~\cite{chun2020multi}, and for the localization of a single dynamic target moving on a plane and tracked by a group of seeker agents~\cite{dou2020target,chen2023target}.
In these works, the seeker agents are steered along continuous predetermined (elliptical and circular, respectively) trajectories, independently of the target localization uncertainty.
Conversely, trajectory optimization is studied in~\cite{he2019trajectory} where a single seeker agent is required to track a target having unknown dynamics. 
Along the same line, in~\cite{xu2020optimal} the trajectories of multiple sensing agents are designed in order to minimize the mean-squared error on the estimation of a single target position.
In this work, the exploited measurements are both angle and time of arrival. 
\medskip

\noindent{\textit{Contributions}} -  As in~\cite{xu2020optimal}, we propose an estimation and control framework to displace a group of seeker agents in order to optimize the accuracy of the estimated position of a target.
We however deal with bearing measurements.
More in detail, the main contributions are twofold:
\begin{itemize}
    \item on the estimation side, we propose a procedure exploiting the popular weighted least square (WLS) methodology  and entailing the iterative computation of the covariance matrix on the estimated target position;
    \item on the control side, we devise a regulation law for the seeker agents based on the active-sensing approach~\cite{varotto2021active}.
\end{itemize}
We validate the outlined framework through an extensive Monte Carlo (MC) simulation campaign. In light of the achieved results, 
we claim that the goodness of the estimation outcomes is also guaranteed by an ad-hoc initialization procedure for the proposed localization algorithm.
\medskip

\noindent{\textit{Paper Structure}} - The rest of the paper is organized as follows. 
The bearing-based target localization task is detailed in Section~\ref{sec:tlt} with special regard to the measurements modeling. 
In Section~\ref{sec:wls} we illustrate the algorithm to estimate the target position in 3D space. 
Then, in Section~\ref{sec:asca}, we focus on the control strategy to reduce the uncertainty of the estimation outcome. 
The performance of the designed estimation and control framework is discussed in Section~\ref{sec:val}. 
Section~\ref{sec:concl} is devoted to the concluding remarks.


\section{TARGET LOCALIZATION TASK}
\label{sec:tlt}

In this work, we focus on the \textit{bearing-based target localization task} consisting in the estimation of the position of an unknown target through the exploitation of a set of noisy bearing measurements collected by a group of seeker agents.

We assume that the target position $\vec{p}_t \in \real^3$ is fixed over time, while any $i$-th seeker agent, $i \in \mathcal{S} = \{1 \ldots n\}, n \geq 2$, is modeled as a point particle having single-integrator dynamics.
Denoting with $\vec{p}_i(t) \in \real^3$ the time-varying position of the $i$-th seeker agent, it thus holds that 
\begin{equation}
    \label{eq:dot_p_i}
    \dot{\vec{p}}_i(t) = \vec{u}_i(t),
\end{equation}
with $\vec{u}_i(t) \in \real^3$ being the control input chosen in order to guarantee that $\vec{p}_i(t) \neq \vec{p}_t, \forall t$.
To ease the notation, hereafter, we drop the time dependency when not strictly necessary.

\begin{figure}[t!]
    \centering
    \includegraphics[width= 1.0\columnwidth]{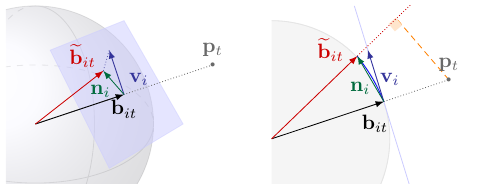}
    \caption{3D (left) and 2D (right) graphical description of the relation between bearing vector $\vec{b}_{it}$ and its noisy measurement $\widetilde{\vec{b}}_{it}$.
    {The blue plane (left) and the blue line (right) represent the tangent spaces $T_{\vec{b}_{it}} \mathbb{S}^2$ and $T_{\vec{b}_{it}} \mathbb{S}^1$ to the 2-sphere and 1-sphere respectively, where the bearing vector $\vec{b}_{it}$ belongs to.}}
    \label{fig:noise_example}
\end{figure}

Any $i$-th seeker agent is supposed to know its position with a certain level of accuracy (e.g. thanks to a high-performance GNSS device)and to be capable of acquiring a noisy measurement of the bearing with respect to the target with sampling period $T>0$ (e.g. through an onboard camera).
Specifically, we define the bearing $\vec{b}_{it}
\in \mathbb{S}^2$ as 
\begin{equation}
    \vec{b}_{it} = \vec{f}_i(\vec{p}_t) = \frac{\vec{p}_t-\vec{p}_i}{\norm{\vec{p}_t-\vec{p}_i}},
\end{equation}
where $\vec{f}_i(\cdot): \real^3 \to \mathbb{S}^2$ is the bearing rigidity function associated to the $i$-th seeker agent~\cite{michieletto2021unified}.
The measurement recorded at any $t=k T$, $k \in \mathbb{N}$, is then modeled as 
\begin{align}
    \label{eq:bearing_meas}
    \widetilde{\vec{b}}_{it} &= \exp_{\vec{b}_{it}} \vec{v}_i = \cos(\norm{\vec{v}_i})\vec{b}_{it} + \frac{\sin(\norm{\vec{v}_i})}{\norm{\vec{v}_i}}\vec{v}_i
\end{align}
{where, for any vector $\vec{b} \in \mathbb{S}^2$, the exponential map $\exp_{\vec{b}}(\cdot) : T_{\vec{b}} \mathbb{S}^2 \to \mathbb{S}^2$ projects onto $\mathbb{S}^2$ the vectors belonging to the tangent space $ T_{\vec{b}} \mathbb{S}^2 $ of $\vec{b}$.} In~\eqref{eq:bearing_meas},  the vector $\vec{v}_i \in T_{\vec{b}_{it}} \mathbb{S}^2 $ represents a perturbation that acts orthogonally to $\vec{b}_{it}$ and takes into account the sensor noise and the measurement uncertainties.
In the following, this is modeled as a normally distributed Gaussian random vector having zero mean and positive semi-definite covariance matrix $\vec{\Sigma}_{\vec{v}_i} \in \mathbb{R}^{3 \times 3}$.
In particular, since $\vec{v}_i \in T_{\vec{b}_{it}} \mathbb{S}^2 \subset \real^3$, the covariance matrix is required to have a zero eigenvalue corresponding to the eigenvector $\vec{b}_{it}$.
To ensure this property, we define $\vec{\Sigma}_{\vec{v}_i}$ as
\begin{equation}
    \label{eq:Sigma_vi}
    \vec{\Sigma}_{\vec{v}_i} = \vec{P}(\vec{b}_{it})\vec{\Sigma}_{\bar{\vec{v}}_i}\vec{P}(\vec{b}_{it})^\top,
\end{equation}
where $\vec{\Sigma}_{\bar{\vec{v}}_i} \in \mathbb{R}^{3 \times 3}$ is a suitably selected positive definite matrix and $\vec{P}(\cdot): \real^3 \to \real^{3 \times 3}$ is the orthogonal projection operator such that $\vec{P}(\vec{x}) = \vec{I}_3 - (\vec{x}^\top \vec{x})/(\vec{x}\vec{x}^\top)$ with $\vec{x} \in \mathbb{R}^3$.
A graphical representation of the bearing measurement~\eqref{eq:bearing_meas} is given in Figure~\ref{fig:noise_example}, highlighting the relation between the bearing $\vec{b}_{it}$ and the perturbation vector $\vec{v}_i$. {The figure schematically reports on the left the general case of a 3D scenario, mainly considered in this work, while on the right it is given, just for the sake of simplicity, also a planar case.}

Finally, we assume that the seeker agents constitute an ideal formation in terms of communication, meaning that each group component can exchange data with all the others without delays, interference, and/or information corruption and degradation.
Accounting for the graph-based representation of the formations, the seeker agents group is thus modeled as a complete indirect graph.


\section{WLS-BASED TARGET POSITION ESTIMATION}
\label{sec:wls}

The noisy bearing measurements introduced in Section~\ref{sec:tlt} can be interpreted according to the popular additive noise model.
Indeed, the expression~\eqref{eq:bearing_meas} of $\widetilde{\vec{b}}_{it}$ can be rewritten as 
\begin{equation}
    \label{eq:bearing_meas_add}
    \widetilde{\vec{b}}_{it} = {\vec{b}}_{it} + \vec{n}_i=\vec{f}_i(\vec{p}_t) + \vec{n}_i,
\end{equation}
where $\vec{n}_i \in \real^3, i \in \mathcal{S},$ is a function of both the bearing $\vec{b}_{it} \in \mathbb{S}^2$ and the perturbation vector $\vec{v}_i \in \mathbb{S}^2$, namely
\begin{equation}
    \label{eq:n_i_explained}
    \vec{n}_i= \left( \cos(\norm{\vec{v}_i}) -1 \right) \vec{b}_{it} + \frac{\sin(\norm{\vec{v}_i})}{\norm{\vec{v}_i}}\vec{v}_i.
\end{equation}
Note that Figure~\ref{fig:noise_example} clarifies also the relation between the vectors $\vec{v}_i \in T_{\vec{b}_{it}} \mathbb{S}^2 \subset \real^3$ and $\vec{n}_i \in \real^3$.

All the seeker agents measurements can then be collected in the \textit{bearing measurements vector} $\widetilde{\vec{b}}_{{S}} \in \mathbb{S}^{2^n}$, defined as 
\begin{equation}
    \label{eq:bearing_meas_vect}
    \widetilde{\vec{b}}_{{S}} = \colvec{
        \widetilde{\vec{b}}_1 \\ \vdots \\ \widetilde{\vec{b}}_n
    } =
    \colvec{
        \vec{f}_1(\vec{p}_t) \\ \vdots \\ \vec{f}_n(\vec{p}_t)
    } + \colvec{
        \vec{n}_1 \\ \vdots \\ \vec{n}_n
    } =
    \vec{f}_{{S}}(\vec{p}_t) + \vec{n}_{{S}},
\end{equation}
where $\vec{f}_{{S}}(\cdot): \real^{3n} \to \mathbb{S}^{2^n}$ is the standard bearing function when the $\mathbb{R}^3$ domain is taken into account~\cite{michieletto2021unified}.

Since any $\vec{v}_i$ in~\eqref{eq:bearing_meas} is modeled as a normally distributed Gaussian random vector, we suppose that also $\vec{n}_{{S}}$ in~\eqref{eq:bearing_meas_vect} approximates a Gaussian random vector with $n$ i.i.d. components.
The soundness of this assumption will be confirmed by the numerical results in Section~\ref{sec:val:cov_val}.
In light of this fact, we cope with the bearing-based target localization task by adopting the well-stated WLS approach.
In doing this, the target position estimate $\hat{\vec{p}}_t \in \mathbb{R}^3$ is determined as 
\begin{equation}
    \label{eq:p_t_hat}
    \hat{\vec{p}}_t = \argmin_{\vec{p} \in \mathbb{R}^3} \norm{\widetilde{\vec{b}}_{{S}} - \vec{f}_{{S}}(\vec{p})}_{\vec{W}}^2,
\end{equation}
where $\vec{W} \in \mathbb{R}^{3n \times 3n}$ is a positive definite weight matrix whose selection is discussed in the following.

\subsection{Iterative WLS Solution}
\label{sec:wls:its}

The WLS problem~\eqref{eq:p_t_hat} can be solved via the iterative procedure described in Algorithm~\ref{alg:ils} wherein, at each iteration, it is computed the solution of the linearized WLS problem based on the linearization of the bearing rigidity function around the current target position estimate (lines 3-5).

Remarkably, the Jacobian matrix $\vec{F}_{{S}}(\cdot) \in \mathbb{R}^{3n \times n}$ of the function $\vec{f}_{{S}}(\cdot)$ can be computed in closed form and it results from the stacking of scaled projection operators, namely
\begin{equation}
    \label{eq:Jf}
    \vec{F}_{S}(\hat{\vec{p}}_t) = 
    \frac{\partial\vec{f}_{S}(\vec{p})}{\partial \vec{p}}\biggr\vert_{\vec{p}=\hat{\vec{p}}_t} =
    \colvec{
        \frac{1}{\norm{\hat{\vec{p}}_t-\vec{p}_1}}\vec{P}( \vec{f}_1 (\hat{\vec{p}}_t)) \\
        \vdots \\
        \frac{1}{\norm{\hat{\vec{p}}_t-\vec{p}_n}} \vec{P}( \vec{f}_n (\hat{\vec{p}}_t))
    }.
\end{equation}
We observe that~\eqref{eq:Jf} corresponds to the bearing rigidity matrix associated with the $n+1$ multi-agent formation embedded in $\real^3$ and composed by $n$ seeker agents and the target. In this case, the formation sensing capabilities are modeled by a directed star graph having the target as the center of the ingoing edges representing the bearing measurements taken by the $n$ seeker agents.

The algorithm stops when $\norm{\Delta\hat{\vec{p}}_t} \in \mathbb{R}^+$, quantifying the innovation on the target position estimate with respect to the previous iteration, is below a given threshold $\epsilon \in \mathbb{R}^+$ (line~6).

\begin{algorithm}[t!]
    \caption{Iterative WLS algorithm.}
    \label{alg:ils}
    \KwData{$\hat{\vec{p}}_{t,0}$, $\widetilde{\vec{b}}_{S}$, $\epsilon$, $\vec{W}$}
    \KwResult{$\hat{\vec{p}}_t$}
    $\hat{\vec{p}}_t \gets \hat{\vec{p}}_{t,0}$
    \;
    \Do{$\norm{\Delta \hat{\vec{p}}_t}>\epsilon$}{
      $\Delta \widetilde{\vec{b}} \gets \widetilde{\vec{b}}_{S} - \vec{F}_{S}(\hat{\vec{p}}_t)$\;
      $\Delta\hat{\vec{p}}_t \gets \left(\vec{F}_{S}(\hat{\vec{p}}_t)^\top \vec{W} \vec{F}_{S}(\hat{\vec{p}}_t) \right)^{-1} \vec{F}_{S}(\hat{\vec{p}}_t)^\top \vec{W}\Delta \widetilde{\vec{b}}$\;
      $\hat{\vec{p}}_t \gets \hat{\vec{p}}_t + \Delta \hat{\vec{p}}_t$\;
    }
\end{algorithm}

\subsection{Algorithm Initialization}
\label{sec:wls:alg_init}

The goodness of the target position estimate obtained from Algorithm~\ref{alg:ils} depends on its initialization (line 1).

If any prior information is available, e.g., in correspondence to the first measurements acquisition, a suitable initial estimate guess $\hat{\vec{p}}_{t,0} \in \mathbb{R}^3$ is the point that minimizes the distance from the lines passing through $\vec{p}_i$ and directed along $\widetilde{\vec{b}}_{it}, \forall i \in \{1 \dots n\}$.
Introducing the seeker agents position vector $\vec{p}_{S}= \colvec{ \vec{p}_1 \ldots \vec{p}_n}^\top \in \mathbb{R}^{3n}$, we have that 
\begin{equation}    \label{eq:line_distance_minimization_problem}
    \hat{\vec{p}}_{t,0} = \argmin_{\vec{p}\in \mathbb{R}^3} d(\vec{p} \mid \vec{p}_{S},\widetilde{\vec{b}}_{S}),
\end{equation}
where $d(\cdot \mid \vec{p}_{S},\widetilde{\vec{b}}_{S}): \mathbb{R}^3 \to \mathbb{R}$ is the cumulative distance function given the seeker agents position and bearing measurements, namely it is
\begin{equation}
    \label{eq:line_cost}
    d(\vec{p} \mid \vec{p}_{S},\widetilde{\vec{b}}_{S}) = \frac{1}{2} \sum_{i=1}^n(\vec{p}_i-\vec{p})^\top \vec{P}(\widetilde{\vec{b}}_{it}) (\vec{p}_i-\vec{p}).
\end{equation}
Observe that function~\eqref{eq:line_cost} is convex and quadratic with respect to the argument $\vec{p}$, thus the global minimum for the problem~\eqref{eq:line_distance_minimization_problem} can be computed in closed form as
\begin{equation}
    \label{eq:phat_0}
    \hat{\vec{p}}_{t,0} = \left(\vec{P}_{S}(\widetilde{\vec{b}}_{S})^\top\vec{P}_{S}(\widetilde{\vec{b}}_{S})\right)^{-1} \vec{P}_{S}(\widetilde{\vec{b}}_{S})
\vec{p}_{S},
\end{equation}
where $\vec{P}_{S}(\cdot): \mathbb{S}^{2^n} \to \real^{3n \times 3}$ is the stacked projectors operator, so that it is
$
\vec{P}_{S}(\widetilde{\vec{b}}_{S}) =
\colvec{
    \vec{P}(\widetilde{\vec{b}}_{1t}) &
    \ldots &
    \vec{P}(\widetilde{\vec{b}}_{nt})
}^\top
$.
Thus, exploiting the idempotent property of $\vec{P}_{S}(\cdot)$, the vector $\hat{\vec{p}}_{t,0}$ can be retrieved as
\begin{equation}
	\label{eq:phat_0_rev}
	\hat{\vec{p}}_{t,0} = \vec{A}^{-1}\vec{y}
\end{equation}
with matrix $\vec{A} \in \mathbb{R}^{3 \times 3}$ and vector $\vec{y} \in \mathbb{R}^3$ defined as
\begin{equation}
    \label{eq:A_and_b}
    \vec{A} = \sum_{i=1}^n \vec{P}(\widetilde{\vec{b}}_{it}) \quad \text{and} \quad \vec{y} = \sum_{i=1}^n \vec{P}(\widetilde{\vec{b}}_{it})\vec{p}_i.
\end{equation}
Specifically, the existence of the solution~\eqref{eq:phat_0_rev} is guaranteed by the full rankness of the matrix $\vec{A}$: this condition holds when at least two bearing measurements are not collinear.

When a target position estimation is already available, e.g., based on some previous measurements set, the Algorithm~\ref{alg:ils} can be initialized by exploiting this information.
In this case, it is, e.g., $\hat{\vec{p}}_{t,0}(t+ T) = \hat{\vec{p}}_t(t)$.

\begin{remark}
Addressing problem~\eqref{eq:line_distance_minimization_problem} implies the computation of a target position estimation, which turns out to be a suboptimal solution for the target localization task formalized as in~\eqref{eq:p_t_hat}.
This is due to the fact that the minimization problem~\eqref{eq:p_t_hat} involves the notion of chordal distance over spherical manifold, 
whereas~\eqref{eq:line_distance_minimization_problem} is based on the line distance.
{With reference to the right panel of} Figure~\ref{fig:noise_example}, we remark that~\eqref{eq:phat_0_rev} is computed by minimizing the projection along the direction of $\widetilde{\vec{b}}_{it}$ (in orange), whereas the solution of~\eqref{eq:p_t_hat} is determined by minimizing the norm of the vector $\vec{n}_i$ (in green).
\end{remark}

\subsection{Localization Uncertainty}
\label{sec:wls:lu}

The covariance matrix associated with the target position estimates $\hat{\vec{p}}_t$ from Algorithm~\ref{alg:ils} can be approximated using linear regression theory.
Accordingly, we have that
\begin{equation} \label{eq:estim_covariance}
\begin{aligned}
    \vec{\Sigma}_{\hat{\vec{p}}_t} = \left( (\vec{F}_{S}(\hat{\vec{p}}_t)^\top \vec{W} \vec{F}_{S}(\hat{\vec{p}}_t))^{-1} \vec{F}_{S}(\hat{\vec{p}}_t)^\top \vec{W} \right) \vec{\Sigma}_{\vec{n}_{S}}& \\ 
    \left( (\vec{F}_{S}(\hat{\vec{p}}_t)^\top \vec{W} \vec{F}_{S}(\hat{\vec{p}}_t))^{-1} \vec{F}_{S}(\hat{\vec{p}}_t)^\top \vec{W} \right)^\top,&
    \end{aligned}
\end{equation}
where $\vec{\Sigma}_{\vec{n}_{S}} \in \mathbb{R}^{3n \times 3n}$ is the positive definite covariance matrix of the noise vector $\vec{n}_{S}$.
The validity of~\eqref{eq:estim_covariance} is confirmed by the numerical results that will be presented in Section~\ref{sec:val:cov_val}.

In the particular case wherein $\Sigma_{\vec{n}_{S}}$ is known with a certain level of confidence, it is suitable to select the weight matrix as $\vec{W} = \vec{\Sigma}_{\vec{n}_{S}}^{-1}$ in order to have
\begin{equation}
    \label{eq:Sigma_p_t_hat}
    \vec{\Sigma}_{\hat{\vec{p}}_t} = \left( \vec{F}_{S}(\hat{\vec{p}}_t)^\top \vec{\Sigma}_{\vec{n}_{S}}^{-1} \vec{F}_{S}(\hat{\vec{p}}_t) \right)^{-1}.
\end{equation}
In this way, the target position estimate covariance matrix $\vec{\Sigma}_{\hat{\vec{p}}_t}$ results to be dependent on the noise covariance matrix $\vec{\Sigma}_{\vec{n}_{S}}$ and on the matrix $\vec{F}_{S}(\hat{\vec{p}}_t)$, which summarizes the relative position between the seeker agents and the target based on its current position estimate.
This fact supports the employment of the active-sensing paradigm in the design of the seeker agents controller: loosely speaking, the idea is to steer the seeker agents in order to minimize the uncertainty on the estimated position of the target.


\section{ACTIVE-SENSING CONTROL APPROACH}
\label{sec:asca}

To develop an ad-hoc active-sensing control approach for the seeker agents group, we rest on the fact that the covariance matrix $\vec{\Sigma}_{\hat{\vec{p}}_t}$ is a function of the seeker agents position $\vec{p}_{S}$ (i.e., it is $\vec{\Sigma}_{\hat{\vec{p}}_t}=\vec{\Sigma}_{\hat{\vec{p}}_t}({\vec{p}}_{S})$) and its determinant constitutes a measure of the volume of the uncertainty ellipsoid associated to the target position estimate $\hat{\vec{p}}_t$.
Thus, given that $\det(\vec{M}^{-1}) = \det(\vec{M})^{-1}$ for any nonsingular matrix $\vec{M}$, we design an active-sensing control law based on the maximization of the estimation reward
\begin{subequations}\label{eq:active_sense_reward}
    \begin{align}
        J({\vec{p}}_{S}) &= \det\left((\vec{\Sigma}_{\hat{\vec{p}}_t}({\vec{p}}_{S}))^{-1}\right) \\
        &= \det(\vec{F}_S(\hat{\vec{p}}_t)^\top \Sigma_{\vec{n}_{S}}^{-1} \vec{F}_S(\hat{\vec{p}}_t)),
        \end{align}
\end{subequations}
relying also upon the weight matrix selection $\vec{W} = \vec{\Sigma}_{\vec{n}_{S}}^{-1}$.

The cost function~\eqref{eq:active_sense_reward} is unbounded from above and it has a singularity when $\vec{p}_i \to  \hat{\vec{p}}_t$ for at least a seeker agent (in this case, it holds that $J(\vec{p}_S) \to +\infty$).
On the other hand, the uncertainty on the target position estimate is intuitively reduced when the seeker agents approach the target itself.

Now, under the requirement of designing the control input $\vec{u}_i$ of any $i$-th seeker agent in order to guarantee that $\vec{p}_i\neq \vec{p}_t$, we address the minimization of the estimation reward~\eqref{eq:active_sense_reward} in a constrained framework, imposing all seeker agents to move while maintaining their estimated distance with respect to the target.
In light of the assumed single integrator dynamic model~\eqref{eq:dot_p_i}, the given constraint can be fulfilled by adopting a \textit{projected gradient ascend control law}.
Formally, for any $i$-th seeker agent, the control input is computed as 
\begin{subequations}
    \label{eq:constr_max_J_u}
    \begin{align}
    &\vec{u}_i = k \; \vec{P}(\hat{\vec{b}}_{it}) \vec{J}(\vec{p}_{S}) \quad \text{with} \\
&\hat{\vec{b}}_{it}=\frac{\hat{\vec{p}}_t-\vec{p}_i}{\norm{\hat{\vec{p}}_t-\vec{p}_i}}, \quad  \vec{J}(\vec{p}_{S})=\frac{\partial J(\vec{p}_S)}{\partial \vec{p}_i},
\end{align}
\end{subequations}
where $k \in \mathbb{R}^+$ is a tunable gain parameter.
We remark that $\hat{\vec{b}}_{it} \in \mathbb{S}^2$ in~\eqref{eq:constr_max_J_u} represents the estimated bearing of the $i$-th seeker agent with respect to the target.
Thus, the projector operator $\vec{P}(\hat{\vec{b}}_{it}) \in \mathbb{R}^{3 \times 3}$ ensures that the $i$-th seeker agent moves on the sphere centered in the estimated target position with radius $\hat{d}_{it} = \norm{\hat{\vec{p}}_t-\vec{p}_i} \in \mathbb{R}$.
We also point out that the Jacobian matrix $\vec{J}(\vec{p}_{S}) \in \mathbb{R}^{3 \times 3}$ in~\eqref{eq:constr_max_J_u} can be computed as the sum of two terms, namely
\begin{equation}
    \label{eq:gradJ_decomposed}
    \vec{J}(\vec{p}_{S}) = \frac{\partial J(\vec{p}_S)}{\partial\hat{\vec{b}}_{it}} \frac{\partial \hat{\vec{b}}_{it}}{\partial \vec{p}_i} + \frac{\partial J(\vec{p}_S)}{\partial\hat{d}_{it}} \frac{\partial \hat{d}_{it}}{\partial \vec{p}_i}.
\end{equation}
The first addendum in~\eqref{eq:gradJ_decomposed} belongs to the column space of the matrix $\vec{P}(\hat{\vec{b}}_{it})$, i.e., to $\image (\vec{P}(\hat{\vec{b}}_{it}))$, while the second addendum is contained in its null space, i.e., $\ker(\vec{P}(\hat{\vec{b}}_{it}))$.
Then, the control input~\eqref{eq:constr_max_J_u} can be equivalently rewritten as
\begin{equation}
    \label{eq:constr_max_J_u2}
    \vec{u}_i = k \; \frac{\partial J(\vec{p}_S)}{\partial\hat{\vec{b}}_{it}} \frac{\partial \hat{\vec{b}}_{it}}{\partial \vec{p}_i}.
\end{equation}

By accounting for the case wherein the noise vectors $\vec{n}_1, \ldots \vec{n}_n$ are uncorrelated and characterized by diagonal covariance matrices, namely when $\vec{\Sigma}_{\vec{n}_S} = \diag(\vec{\Sigma}_{\vec{n}_1}, \dots \vec{\Sigma}_{\vec{n}_n})$ and $\vec{\Sigma}_{\vec{n}_i}= \sigma_i^2 \vec{I}_3$ for all $i \in \mathcal{S}$, it is possible to express the control input~\eqref{eq:constr_max_J_u} in a more convenient closed form.
Under these conditions, indeed, the matrix $\vec{\Sigma}_{\hat{\vec{p}}_t}^{-1}$ results to be 
\begin{equation} \label{eq:Sigma_dec}
        \vec{\Sigma}_{\hat{\vec{p}}_t}^{-1}
        = \sum_{i=1}^n \frac{1}{\sigma_i^{2}} \frac{1}{\hat{d}_{it}^2}\vec{P}(\hat{\vec{b}}_{it}),
\end{equation}
and it can be interpreted as a weighted version of the matrix $\vec{A}$ in~\eqref{eq:A_and_b}.
Note also that $\hat{d}_{it}$ and $\hat{\vec{b}}_{it}$ in~\eqref{eq:Sigma_dec} can be seen as independent variables that account for the radial and tangent direction of the gradient of the estimation reward~\eqref{eq:active_sense_reward}.
Hence, $J(\vec{p}_S)$ is trivially maximized by reducing the estimated distance of the seeker agents with respect to the target or by relocating the seeker agents.
This observation is supported by the numerical results provided in Section~\ref{sec:val:cov_val}.


\begin{figure*}[t!]
    \begin{subfigure}[b]{0.33\textwidth}
        \centering
        \includegraphics[height=4cm]{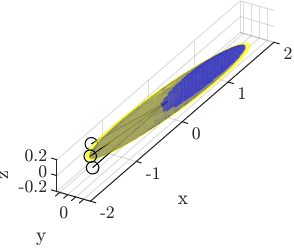}
        \caption{clustered seeker agents configuration \\[0.1cm] 
                \begin{tabular}{l}
        $\vec{p}_S = \colvec[0.8]{-15 \;\; 0 \;\; 0 & -15 \;\; 3 \;\; 0 & -15 \;\; 0 \;\; 1 }$ \si{\metre} \\
        $\theta_M = \SI{0.16}{\radian}$\\[0.1cm]
                    $\bar{\hat{\vec{p}}}_t = \colvec[0.8]{ 0.12 \;\; 0 \;\; 0}$ \si{\metre}  \\[0.1cm]
        $J({\vec{p}}_{S}) = \SI{0.755e3}{} $, \;
        $c_\Sigma = 105$
        \end{tabular}}
        \label{fig:cov:bad}
    \end{subfigure}
    \begin{subfigure}[b]{0.33\textwidth}
        \centering
        \includegraphics[height=4cm]{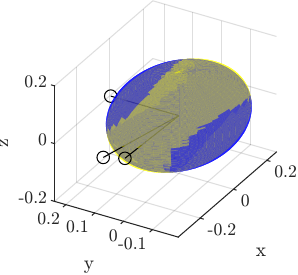}
        \caption{spaced seeker agents configuration
        \\[0.1cm] 
                \begin{tabular}{l}
        $\vec{p}_S = \colvec[0.8]{0 \;\; 15 \;\; 0 & -15 \;\; 3 \;\; 0 & -15 \;\; 0 \;\; 1 }$ \si{\metre} \\
        $\theta_M = \SI{1.05}{\radian}$\\[0.1cm]
                    $\bar{\hat{\vec{p}}}_t = \colvec[0.8]{ 0 \;\; 0 \;\; 0}$ \si{\metre}  \\[0.1cm]
        $J({\vec{p}}_{S}) = \SI{18e3}{} $, \;
        $c_\Sigma =3$
        \end{tabular}}
        \label{fig:cov:good}
    \end{subfigure}
    \begin{subfigure}[b]{0.33\textwidth}
        \centering
        \includegraphics[height=4cm]{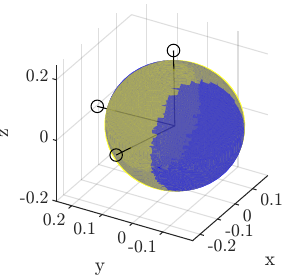}
        \caption{well-spaced seeker agents configuration\\[0.1cm] 
                \begin{tabular}{l}
        $\vec{p}_S = \colvec[0.8]{0 \;\; 15 \;\; 0 & -15 \;\; 3 \;\; 0 & 0 \;\; 0 \;\; 15 }$ \si{\metre} \\
        $\theta_M = \SI{1.47}{\radian}$\\[0.1cm]
                    $\bar{\hat{\vec{p}}}_t = \colvec[0.8]{ 0 \;\; 0 \;\; 0}$ \si{\metre}  \\[0.1cm]
        $J({\vec{p}}_{S}) = \SI{24e3}{} $, \;
        $c_\Sigma =1.2$
        \end{tabular}}
        \label{fig:cov:best}
    \end{subfigure}
    \caption{Ellipsoids representing the theoretical $\vec{\Sigma}_{\hat{\vec{p}}_t}$ (yellow) and the empirical $\hat{\vec{\Sigma}}_{\hat{\vec{p}}_t}$ (blue) covariance matrix in correspondence to different seeker agents configurations. 
    The bearing measurements are marked in black. Note that different ranges have been employed for the figure axes.}
    \label{fig:covariances validation}
\end{figure*}
\section{NUMERICAL VALIDATION}
\label{sec:val}

We first assess the validity of the approximation~\eqref{eq:estim_covariance}
of the target position estimate covariance matrix.
Then we investigate the performance of the outlined WLS-based target position observer and the designed active-sensing controller.

\subsection{Model Validation: Localization Uncertainty}
\label{sec:val:cov_val}

The expression~\eqref{eq:estim_covariance} of the target position estimate covariance matrix comes from the assumption that $\vec{n}_S$ approximates a Gaussian random vector having zero mean and a certain covariance matrix $\vec{\Sigma}_{\vec{n}_S}$.
According to~\eqref{eq:n_i_explained}, any $i$-th component $\vec{n}_i, i \in \mathcal{S}$ of the vector $\vec{n}_{S}$ results from the sum of two orthogonal terms.
The first term is a downscaled version of the vector $\vec{v}_i$, hence the Gaussian-based modeling turns out to be fairly adequate, while this is not valid for the second term of~\eqref{eq:n_i_explained} which entails the corresponding measurement $\widetilde{\vec{b}}_{it}$ to belong to the unit sphere $\mathbb{S}^2$.

Given these premises,  we remark that the performance of Algorithm~\ref{alg:ils} rests upon the selection $\vec{W} = \vec{\Sigma}_{\vec{n}_{S}}^{-1}$, hence it is necessary to ensure a good knowledge of the covariance matrix of the vector $\vec{n}_S$ whose components are hereafter supposed to be uncorrelated and characterized by diagonal covariance matrices.
In the following, thus, we focus on the effectiveness of the assumed approximation
 \begin{equation}
     \label{eq:cov_eq}
\vec{\Sigma}_{\vec{n}_{S}} \approx \vec{\Sigma}_{\bar{\vec{v}}_S}
\end{equation}
where $\vec{\Sigma}_{\bar{\vec{v}}_S}= \diag\left( \vec{\Sigma}_{\bar{\vec{v}}_i} \right) \in \mathbb{R}^{3n \times 3n}$ being $\vec{\Sigma}_{\bar{\vec{v}}_i}$ the design parameter introduced in~\eqref{eq:Sigma_vi}.

To assess~\eqref{eq:cov_eq}, we perform $N=1000$ tests simulating the target localization task given different bearing noise realizations in correspondence to three different (initial) displacements of $n=3$ seeker agents.
Hereafter, we denote with $S_a, S_b$ and $S_c$ the three considered scenarios, ranging from a more clustered to a well-spaced configuration, and we use the average angle $\theta_M \in [0,\pi]$ between each pair of bearings as an index of the closeness of the seeker agents group.
Formally, this is
\begin{equation}
    \label{eq:closeness}
    \theta_M = \frac{1}{n(n-1)/2}\sum_{i,j \in \mathcal{S}, i \ne j} \arcsin(\norm{\vec{b}_{it} \otimes \vec{b}_{jt}}).
\end{equation}
In all the conducted tests, the target position is always equal to $\vec{p}_t= \colvec{0 \; 0 \; 0}^\top \si{\meter}$.
Moreover, any bearing measurement is generated according to~\eqref{eq:bearing_meas}.
Specifically, the covariance matrix of the corresponding perturbation $\vec{v}_i$ is computed by exploiting~\eqref{eq:Sigma_vi} and setting $\vec{\Sigma}_{\bar{\vec{v}}_i} = (\pi/180)\, \vec{I}_3$.
 
We consider the collection $\left \{ \hat{\vec{p}}_{t,k} \right \}_{k=1}^N$ of the (independent) estimates of the target position outputted from Algorithm~\ref{alg:ils}, whose stopping condition is regulated by the parameter $\epsilon=\SI{e-4}{\metre}$.
Then, we determine the empirical target position estimate covariance matrix $\hat{\vec{\Sigma}}_{\hat{\vec{p}}_t} \in \mathbb{R}^{3 \times 3}$ as
\begin{equation}
    \label{eq:emp_mean}
    \hat{\vec{\Sigma}}_{\hat{\vec{p}}_t} = \frac{1}{N-1}\sum_{k=1}^N (\hat{\vec{p}}_{t,k} - \bar{\hat{\vec{p}}}_t)(\hat{\vec{p}}_{t,k} - \bar{\hat{\vec{p}}}_t)^\top,
\end{equation}
where $\bar{\hat{\vec{p}}}_t \in \mathbb{R}^3$ indicates the empirical target position estimate mean, namely it is $\bar{\hat{\vec{p}}}_t= 1/N \sum_{k=1}^N \hat{\vec{p}}_{t,k}$.

Figure~\ref{fig:covariances validation} clears up the relation between the covariance matrix ${\vec{\Sigma}}_{\hat{\vec{p}}_t}$ computed as in~\eqref{eq:Sigma_p_t_hat} (hereafter termed \textit{theoretical covariance}) and its empirical counterpart~\eqref{eq:emp_mean} (hereafter termed \textit{empirical covariance}), by reporting the ellipsoidal covariance representation in correspondence to the $S_a, S_b$ and $S_c$ configurations.
The results in Figure~\ref{fig:covariances validation} lead to two main observations.
First, the theoretical covariance over-approximates the empirical one, ensuring a conservative target position estimation.
Second, the difference between the two covariances reduces when the seeker agents are more spread (see Fig.~\ref{fig:cov:best} as compared to Fig.~\ref{fig:cov:bad}).
This last point will be better discussed in the rest of the section; nonetheless, we can conclude that since the theoretical covariance is a good approximation of the empirical, especially in the more probable case in which the seeker agents are not all clustered together, the soundness of assumption~\eqref{eq:cov_eq} is guaranteed.

\subsection{Solution Validation: Observer Performance}
\label{sec:val:obs}

The tests described in the previous section provide also some insights into the performance of the proposed target position estimation method, i.e., of Algorithm~\ref{alg:ils}.
Indeed, the empirical mean $\bar{\hat{\vec{p}}}_t$ of the estimated target position turns out to be biased for clustered configurations of seeker agents ($\bar{\hat{\vec{p}}}_t = \colvec{ 0.12 \; 0 \; 0}$ in correspondence to $S_a$ configuration).
In addition, the clustered seeker agents configuration leads to a high volume of uncertainty and also a low estimation reward, while the well-spaced configuration $S_c$ ensures greater accuracy.
This is justified by the fact that the volume of the covariance ellipsoids strictly depends on the position $\vec{p}_S$ of the seeker agents, but also on their relative position with respect to the target according to~\eqref{eq:Sigma_dec}.
In the three considered scenarios, the seeker agents are roughly placed at $\SI{15}{\metre}$ from the target but the direction of the bearings varies.
Intuitively, any bearing identifies an infinite set of target position estimates lying on the line passing through the corresponding seeker agent position.
To resolve the correct distance, thus, at least another not collinear bearing is needed and, in particular, the more the second bearing is orthogonal to the first one, the greater the information provided.
Lastly, we observe that the well-spaced seeker agents configuration $S_c$ yields also a covariance ellipsoid less flattened at the poles: in this case, the condition number of the theoretical covariance $c_\Sigma \in \mathbb{R}$ approaches its minimum value.
Note that when $c_\Sigma=1$ the covariance ellipsoid is a sphere, in this case, the localization uncertainty is favorably uniformly distributed on the three position components. 

\begin{figure}[t]
    \centering
    \includegraphics[width=0.9\columnwidth]{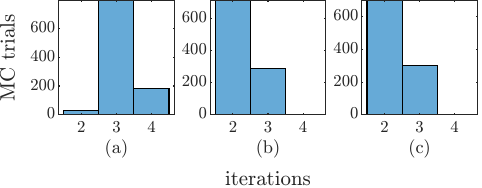}
    \caption{Number of iterations performed by the Algorithm~\ref{alg:ils}
    in $N=1000$ MC trials considering different seeker agents configurations. }
    \label{fig:hist}
\end{figure}

Figure~\ref{fig:hist} shows the distribution of the MC trials over the number of iterations required by Algorithm~\ref{alg:ils} to reach the stopping condition: it results that this value  
is no greater than four in all the considered scenarios.
This confirms the goodness of the adopted initialization strategy (Sec.~\ref{sec:wls:its}).

\subsection{Solution Validation: Controller Performance}

To investigate the performance of the outlined active-sensing control approach, we take into account the scenario wherein the target position estimation poorly performs.
Thus, the attention is focused on the effects of the application of the control law~\eqref{eq:constr_max_J_u2} to any seeker agent composing the clustered configuration $S_a$.
In doing this, we set the bearing measurements acquisition period to $T=\SI{0.1}{\second}$ and the control gain to $k=0.002$.

\begin{figure}[t!]
    \centering
    \includegraphics[width=0.98\columnwidth]{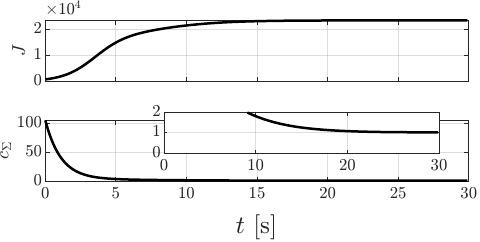}
    \caption{Estimation reward and condition number of the (theoretical) covariance matrix, while controlling the seeker agents configuration $S_a$.}
    \label{fig:contr_perf}
\end{figure}

\begin{figure}[t!]
    \centering
    \includegraphics[width=0.8\columnwidth]{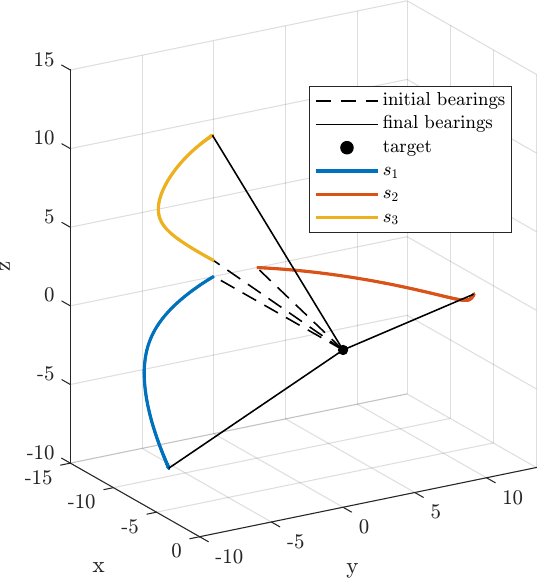}
    \caption{Application of the proposed active-sensing control approach to the initial seeker agent configuration $S_a$;  blue, yellow, and orange lines represent the agents trajectories.}
    \label{fig:seekers_trj}
\end{figure}

Figure~\ref{fig:contr_perf} reports the trend of the estimation reward $J(\vec{p}_S)$ and of the condition number of the covariance matrix $\vec{\Sigma}_{\hat{\vec{p}}_t}$
Note that $J(\vec{p}_S)$ reaches the steady-state value $\sim$\num{23e3} in almost $\SI{15}{s}$, concurrently the condition number converges to \num{1.02} meaning that the covariance ellipsoid approximates a sphere.
The trajectories of the seeker agents $s_1, s_2, s_3$ are finally depicted in Figure~\ref{fig:seekers_trj}: starting from the clustered configuration $S_a$, the seeker agents are steered in some new positions such that the resulting bearings (solid black lines) are almost orthogonal among them.
We observe also that the proposed control law~\eqref{eq:constr_max_J_u} should guarantee that the seeker agents maintain the constant distance from the current target estimation.
This is not ensured in the practical discrete implementation, however, such a distance can not decrease since any seeker agent velocity vector turns out to be tangential to the bearing with respect to the target.

\section{CONCLUSIONS}
\label{sec:concl}

In this work, we cope with the localization of a static target by a group of seeker agents having bearing sensing capabilities.
The proposed solution involves both the target position estimate and the seeker agents control.
The former is faced through an iterative WLS algorithm whose ad-hoc initialization procedure improves the convergence.
The latter is tackled in the active-sensing framework by resting on a projected gradient descent technique.
The two issues are linked by the covariance on the target position which is computed in the estimation phase and optimized in the control one. 
The performance of the proposed solution is investigated via numerical simulations whose results highlight the role of the distance and the relative position between sensing agents and the target.
Specifically, the accuracy of target position estimate is improved by minimizing the distance of the seeker agents or by orthogonalizing their recorded bearing measurements. 

Motivated by these encouraging results, in the future, we intend to study the case of dynamic targets, as well as, exploit the bearing rigidity theory to cope with partial sensing data.


\bibliographystyle{IEEEtran}
\bibliography{L-CSS23}

\end{document}